\documentstyle[12pt]{article}
\begin{document}

\begin{center}
{\bf SHOCK-LIKE FREEZE-OUT IN RELATIVISTIC \\
HYDRODYNAMICS}

\vspace{1.0cm}

{\bf Kyrill A. Bugaev \footnote{Supported by Deutsche
Forschungsgemeinschaft under Contract No. Sa 247/14-1}
} \\

\vspace{1.cm}

Institute for Theoretical Physics\\
University of Hannover\\
30159 Hannover, Germany\\
e-mail: bugaev@itp.uni-hannover.de\\
\end{center}

\vspace{2.0cm}

\begin{abstract}
We have formulated a self-consistent model of freeze-out
on an arbitrary hypersurface. It conserves
energy and momentum across the discontinuity between ideal fluid
and the gas of free particles. Energy and momentum of those free particles
have non-equilibrium values that could be a signal for the
formation of hot and dense matter in heavy ion collisions.
\end{abstract}

\vspace{4.cm}


\newpage
\noindent
{\bf 1. Introduction}
\vspace{0.2cm}

In this paper we present our model of freeze-out for relativistic
hydrodynamics. It solves the problem where and how to stop
solving the hydrodynamical equations in such a way that it is consistent
with the conservation of energy and momentum. When one
solves the hydrodynamical equations one has to terminate the solution
on the boundary with the vacuum. Also, one has to stop solving
hydrodynamics at some freeze-out temperature or baryonic density.
However, in most cases one continues to solve the equations for such
values of these quantities, for which hydrodynamics is no longer valid
(see \cite{8,9}).
There are several papers \cite{3,10,11} on a shock-like treatment of the
freeze-out problem. However, in our opinion these approaches are {\em ad hoc},
since the existence of such a shock is postulated and not obtained
as a result of the equations of motion.

It should be noted that this is only part of the trouble. Another one
comes from the problem of calculating particle spectra on a time-like
hypersurface. For a space-like hypersurface of freeze-out
the correct answer for the spectra of particles is given by the
formula of Cooper and Frye \cite{1}. However, one cannot use that formula
for time-like hypersurfaces, since it leads to negative
numbers of particles. This is due to the fact that it was obtained
only for the space-like case, where the decay of one element of gas
does not affect the decay of adjacent elements.

Here we present our model which is based on the conservation laws of
energy and momentum between fluid and gas of free particles. The paper is
organized as follows: the second part contains a short derivation of the
momentum spectra for the gas of free particles on a time-like hypersurface,
which was also obtained recently in \cite{6}; the third
part contains the derivation of the conservation laws; some useful
formulae are derived there; and in the final part we discuss possible
effects on observables.

We also hope that the model suggested here is not only of academic interest,
but will be used by other researchers to solve the equations
of relativistic hydrodynamics in practical calculations.

\vspace{1.cm}

\noindent
{\bf 2. Decay of the gas of free particles}
\vspace{0.2cm}

In order to obtain the particle spectra for the gas, we will use the
method derived by Gorenstein and Sinyukov \cite{2}.
In this paper we shall deduce the equations for the case of
one-dimensional hydrodynamical motion,
but the final result will not depend on this assumption,
since it will be written in covariant form.
Suppose there is a boundary between fluid and gas.
Let us consider the decay of a small element $\Delta x$ of the gas of
free particles in its rest frame. The gas is supposed to be located
in the left hemisphere and to have the
freeze-out temperature $T = T^*$.
(We suppose that the derivative to the freeze-out
hypersurface $v_\sigma$ in the $t-x$ plane is positive). We note that
this frame is the rest frame of the gas {\em before\/} decay. Hereafter
we shall call it the reference frame of the gas.
Suppose the particles in the gas have an equilibrium distribution function
$\phi\left(\frac{p_0}{T^*}\right)$.

First we consider the contribution from particles with negative
momenta that leave the element $\Delta	x$ (see Fig.1)

\begin{equation}
\frac{d N_1}{d^2 p_{\perp} \Delta S_{\perp}} =
\phi\left(\frac{p_0}{T^*}\right) \Delta x\, dp\, \Theta(-p)\,\,\, ,
\end{equation}

\noindent
where $p_{\perp}$ is the transverse momentum of the particle, and
$\Delta S_{\perp}$ the transverse size of the element.
The second contribution is given by particles with negative momenta
from the element $-\frac{p}{p_0} \Delta t$

\begin{equation}
\frac{d N_2}{d^2 p_{\perp} \Delta S_{\perp}} =
- \phi\left(\frac{p_0}{T^*}\right) \frac{p}{p_0}\,
\Delta t \, dp \, \Theta(-p)\,\, .
\end{equation}

\noindent
Finally, the third contribution comes from particles with positive momenta
from the element $\Delta x -\frac{p}{p_0} \Delta t$. However, those
particles will cross the freeze-out hypersurface only if their velocity is
smaller than the derivative to the hypersurface $v_\sigma$ in the $t-x$ plane.
Thus, the third term reads as follows (see Fig.2):

\begin{equation}
\frac{d N_3}{d^2 p_{\perp} \Delta S_{\perp}} =
\phi\left(\frac{p_0}{T^*}\right) \left[\Delta x - \frac{p}{p_0}
\Delta t \right] \, dp \, \Theta(p) \, \Theta\left(v_\sigma - \frac{p}{p_0}
\right)\,\, .
\end{equation}

\noindent
After some simple algebra one obtains the formula for the spectrum of the
gas of free particles

\begin{equation}
\frac{d N_{tot}}{d^2 p_{\perp} d p \Delta S_{\perp}} =
\phi\left(\frac{p_0}{T^*}\right) \left[\Delta x - \frac{p}{p_0}
\Delta t \right] \Theta\left(v_\sigma - \frac{p}{p_0}\right)\,\, .
\end{equation}

\noindent
As will be shown below,
the modification of the spectrum due to the $\Theta$-function
will lead to an energy-momentum tensor that differs from the equilibrium case.

Our last step is to write the formula for the spectrum in a fully relativistic
form. For that we need to change the energy of the particles appearing in the
distribution function in the reference frame of the gas
to the product of the four-vectors of momentum and hydrodynamical
velocity, $p_\mu u^\mu$, and
change the integration over the hypersurface of freeze-out to the product
of the four-vectors of momentum and normal vector
to the freeze-out hypersurface, $p_\mu d\sigma^\mu$. Finally, we have

\begin{equation}
p_0 \frac{d N_{tot}}{d^3 p \Delta S_{\perp}} =
\phi\left(\frac{p_\mu u^\mu}{T^*}\right) p_\nu d \sigma^\nu\,
\Theta\left(p_\rho d \sigma^\rho \right)\,\, ,
\end{equation}

\noindent
where the vector $d \sigma_\mu = (v_\sigma,- 1) dt$ is the normal vector to
the freeze-out hypersurface in the left hemisphere.
It is, however, easy to check that
the above formula is valid for the right hemisphere as well.

The meaning of this nice result is that it is the formula of Cooper and
Frye \cite{1}, but without negative particle numbers!

It is easy to see that for a space-like hypersurface, where $v_\sigma > 1$,
the above expression gives the result obtained by Cooper and Frye \cite{1}.

The energy-momentum tensor of free
particles in the reference frame of the gas has the following form:

\begin{equation}
T^{\mu\nu}_2(v_\sigma) = \int \frac{d^3 p}{p_0}
p^\mu p^\nu \Theta\left(v_\sigma - \frac{p}{p_0}\right)
\phi\left(\frac{p_0}{T}\right)\,\, .
\end{equation}

\noindent
It is easy to calculate this tensor for the case
of noninteracting massless particles; it
has the form:

\begin{eqnarray}
T^{00}_2(v_\sigma) = \epsilon\left(T^*\right) \frac{1 + v_\sigma}{2}\,\, , \\
T^{01}_2(v_\sigma) = \epsilon\left(T^*\right)  \frac{v_\sigma^2 - 1}{4}\,\, ,
 \\
T^{11}_2(v_\sigma) = \epsilon\left(T^*\right)  \frac{v_\sigma^3 + 1}{6}\,\, ,
\end{eqnarray}

\noindent
where $\epsilon$ is the usual energy density. The above result is valid for
the left hemisphere. The corresponding formulae
for the right hemisphere can be obtained in the same way.
Now one can see that for the case
$v_\sigma = 1$ the above expressions give the usual formula for
the ideal gas.
We hope that this unusual behavior of energy and momentum
can be found in experimental pion spectra.

In the paper by Sinyukov \cite{3} another model
for freeze-out was suggested.
He considered the decay from a box and did not account for the additional
contributions from the intrinsic volume of the gas, namely from the element
$-\frac{p}{p_0} \Delta t$. Due to that, the energy-momentum tensor obtained
in his paper is not symmetric! It means that the orbital momentum of the
considered system is not conserved!

On the other hand, he considered an {\em ad hoc\/} model for deflagration from
very hot pionic matter into gas of free particles.
We do not think that such a simple
picture corresponds to the real situation in heavy ion collisions,
since the solution of the hydrodynamical equations \cite{8,9,5}
does not exhibit shock-like transitions in the expansion of hot and dense
pionic matter.

Now we would like to clarify an important question: "What is the difference
between an ideal fluid and the gas of free particles at freeze-out
temperature?"
It seems that the main difference is that they have different values of
the cross-section. Due to that, there are collisions in the fluid which
lead to thermodynamical equilibrium. In contrast, there are about no collisions
in the gas of free particles (and we shall neglect them completely), because the
 cross-section for collisions
is very
small. Of course, fluid and gas have somewhat different values of temperature,
but due to the fact that the mean free-path strongly depends on the temperature,
this difference should be small.
It was found \cite{4} that for pions the mean free-path depends on the
fifth power of the inverse temperature: $ \lambda \approx \frac{const}{T^5} $.
Thus, the difference between the temperatures of the fluid and the gas
should be small, but
due to the strong dependence of the collision cross-section on the
temperature, their mean free-paths should be very different.

Now we would like to formulate a more physical concept for freeze-out,
based on the conservation laws on the discontinuity
between ideal fluid and the gas of free particles.

\vspace{1cm}

\noindent
{\bf 3. Conservation laws on the surface between fluid and gas}
\vspace{0.2cm}

\noindent
We would like to consider a simple model of freeze-out assuming
that the system consists of the fluid and the gas of free particles.
We do not take into account the transition region between them.
Then, the  total energy-momentum tensor of the system is as follows

\begin{equation}
T^{\mu\nu} = T^{\mu\nu}_1 \Theta(T - T_1^*) + T^{\mu\nu}_2
\Theta(T_2^* - T)\,\, ,
\end{equation}

\noindent
where index 1 corresponds to the fluid, and index 2 to the
gas of free particles,
$T$ is the temperature, $T^*_1$ and $T^*_2$ are the freeze-out temperatures
for fluid and gas, respectively.
We assume that those temperatures satisfy $T_1^* \geq T_2^*$.

\noindent
The equations of motion have the form:

\begin{equation}
\partial_\mu T^{\mu\nu} = 0\,\, .
\end{equation}

\noindent
Taking derivatives and using the equations for the evolution of fluid and gas,

\begin{equation}
\partial_\mu T^{\mu\nu}_a = 0\,\, , \,\,\, a = 1, 2\,\, ,
\end{equation}

\noindent
one finds that the terms with the delta-functions must vanish:

\begin{equation}
T^{\mu\nu}_1 \partial_\mu T_1^* = T^{\mu\nu}_2 \partial_\mu T_2^*\,\, .
\end{equation}

Thus, we have obtained two equations (for the 1+1-dimensional case,
but the case of 3+1 dimensions gives the same result).
However, we have to add to this equation the condition that
the derivatives of the temperatures of fluid and gas are equal,
since there is only one hypersurface of freeze-out:

\begin{equation}
\partial_\mu T_1^* = \partial_\mu T_2^*\,\, .
\end{equation}

Using this condition and dividing the first equation above by
the second one and canceling derivatives of $T^*$, we get
a new equation:

\begin{equation}
\left(T^{11}_1 - T^{11}_2 \right)\left(T^{00}_1 - T^{00}_2 \right) =
\left(T^{10}_1 - T^{10}_2 \right)\left(T^{01}_1 - T^{01}_2 \right)\,\, ,
\end{equation}

\noindent
or

\begin{equation}
Det \left(T^{\mu\nu}_1 - T^{\mu\nu}_2 \right) = 0 \,\, .
\end{equation}

However, there is one problem. It seems that we know the energy-momentum
tensor of the gas only in its reference frame, but the
energy-momentum tensor of the
fluid is known in the laboratory system. On the other hand, we do not
know the velocity of the gas of free particles in the laboratory system.
Fortunately, this problem can be easily solved.

Let us suppose that the fluid with temperature $T = T_1^*$ has a velocity
$v_1$ in the laboratory system, and the gas has a velocity
$v_2$ in the same system.
Then, the velocity of the fluid in the reference frame of the
gas is

\begin{equation}
v_{rel} = \frac{v_1 - v_2}{1 - v_1 v_2}\,\, .
\end{equation}

\noindent
Now the energy-momentum tensors of the fluid
$T^{\mu\nu}_1(v_{rel}, T_1^*)$ and the gas
$T^{\mu\nu}_2(v_\sigma, T_2^*)$ are known in
the reference frame of the gas.
And we have one equation that was obtained before:

\begin{equation}
Det \left(T^{\mu\nu}_1(v_{rel},T_1^*) - T^{\mu\nu}_2(v_\sigma,T_2^*)
\right) = 0\,\, .
\end{equation}

\noindent
Fortunately, we have one more equation. Due to the definition of the freeze-out
hypersurface:

\begin{equation}
T(x,t) = T_2^*
\end{equation}

\noindent
and the fact that the velocity
$v_\sigma$ is the derivative of the implicit function  $T(x,t) = T_2^* $
one gets the following equation

\begin{equation}
v_\sigma  =  \left(\frac{d x}{d t} \right)_{T = T_2^*} =
- \frac{\partial_0 T_2^*}{\partial_1 T_2^*} =
\frac{T^{10}_1(v_{rel},T_1^*) - T^{10}_2(v_\sigma,T_2^*)}
{T^{00}_1(v_{rel},T_1^*) - T^{00}_2(v_\sigma,T_2^*)} \,\, .
\end{equation}

\noindent
Another way to obtain the above result is as follows. One has to remember that
the derivatives of the freeze-out temperature $T_2^*$ are the components of the
normal line to the hypersurface $T(x,t) = T_2^*$.
Then, energy-momentum conservation reads

\begin{eqnarray}
T^{11}_1(v_{rel},T_1^*) - T^{10}_1(v_{rel},T_1^*)\, v_\sigma =
T^{11}_2(v_{\sigma},T_2^*) - T^{10}_2(v_{\sigma},T_2^*)\, v_\sigma\,\, , \\
T^{01}_1(v_{rel},T_1^*) - T^{00}_1(v_{rel},T_1^*)\, v_\sigma =
T^{01}_2(v_{\sigma},T_2^*) - T^{00}_2(v_{\sigma},T_2^*)\, v_\sigma \,\, .
\end{eqnarray}

\noindent
Thus, we have two equations for three unknowns $T_1^*$,
$v_{rel}$ and  $v_\sigma$, since we suppose that the freeze-out temperature
for the gas of free particles is known.

From the above equations one finds the freeze-out hypersurface in the reference
frame of the gas. However, this hypersurface is still
unknown in the laboratory frame, because the above equations do not fix the
velocity of the gas in the laboratory frame.
In order to find the freeze-out hypersurface
in the laboratory frame, one has to solve the hydrodynamical equations.
Thus, the problem is solved.

We have to add the following. The equations

\begin{equation}
T^{\mu\nu}_1 \partial_\mu T_2^* = T^{\mu\nu}_2 \partial_\mu T_2^*
\end{equation}

\noindent
are the conservation laws of energy and momentum on the discontinuity
between the fluid and the gas of free particles, since the derivatives of
the freeze-out temperature $T_1^*$ give the normal vector
to the hypersurface. However, one cannot obtain them from the solution
of the equations of hydrodynamics for the fluid. Suppose one knows
the freeze-out point $T(x,t) = T_1^*$ at a fixed time.
In order to find this
point at the next moment of time, one has to solve the hydrodynamical
equations for unphysical temperatures, namely, $T < T_1^*$! Thus,
from the above equations one has to find the velocities $v_\sigma$ and
$v_{rel}$ in the reference frame of the gas. Since the velocity of the
fluid in the laboratory system $v_1$ is known from the solution of the
hydrodynamical
equations, one makes a Lorenz transformation and finds the velocity of
the gas in the laboratory system:

\begin{equation}
v_2 = \frac{v_1 - v_{rel}}{1 - v_1 v_{rel}}\,\, .
\end{equation}

\noindent
Using this velocity one can find the derivative of the freeze-out
hypersurface in the laboratory frame by a Lorenz transformation.

In the case of baryon-rich matter, the freeze-out condition is as follows

\begin{equation}
 n(x,t) = n^*\,\, ,
\end{equation}

\noindent
and the thermodynamical functions depend on temperature and chemical
potential.
In this case we have one more hydrodynamical equation which is
the conservation law of baryonic charge, and one more equation
for the baryonic densities
between fluid and the gas of free particles.

\vspace{1.cm}

\noindent
{\bf 4. Results and discussion}
\vspace{0.2cm}

We have developed a freeze-out model for relativistic hydrodynamics.
We have obtained the conservation laws on the discontinuity between
ideal fluid and the gas of free particles.
We have derived an expression for the energy-momentum
tensor of the gas of free particles and its momentum distribution function.
It is important to emphasize that this tensor differs from
the equilibrium one. Thus, one can hope to find this unusual
behavior in experiments, and
this can give detailed information
about the freeze-out process in heavy ion collisions.

One possible effect is related to measuring the ratio of the total
(longitudinal) momentum density
in the backward hemisphere to the zero component
of the baryonic charge flux. In the reference frame of the gas,
that component reads (for the left hemisphere)

\begin{equation}
N^{0} =  n(T,\mu) \frac{1 + v_\sigma}{2}\,\, .
\end{equation}

\noindent
If it is possible to neglect the contribution of the space-like
part of the freeze-out hypersurface, then that ratio will depend on
the thermodynamical quantities and an average value of $v_\sigma$ on
the time-like part of the freeze-out hypersurface.
Thus, measuring this ratio for the gas of free particles one can
find the average value of $v_\sigma$.
We hope that it is possible to do so for asymmetric heavy ion collisions.
For example, the S + Au reaction that was studied in \cite{7}
indicates that the space-like part of the freeze-out hypersurface is
small and that the velocities on it are approximately constant.
Therefore, finding the suggested ratio can
give important information about the freeze-out process of
hot and dense hadronic matter.

\vspace{1.cm}

\noindent
{\bf Acknowledgment}

\vspace{0.2cm}

The author acknowledges helpful discussions with Prof.\ M.I.\ Gorenstein
and Prof.\ P.U.\ Sauer, and Dr. \ D.H. Rischke for a critical reading of
the manuscript and important comments.

\vspace{1.cm}

\noindent
{\bf Appendix A}

\vspace{0.2cm}

In this Appendix we would like to discuss the question: How to solve the
hydrodynamical equations together with the above equations on the
freeze-out hypersurface.
In fact we shall only show that the suggested system of equations
for the shock is
consistent with the hydrodynamical equations, so that they can be
solved simultaneously.

\noindent
{\bf A.} Suppose our system consists of pions only.
In this case thermodynamical quantities
depend only on the temperature.
Suppose we have two hydrodynamical equations for the fluid that can
be written in the form

\begin{equation}
F_a \left(\epsilon_{1i}, p_{1i}, v_{1i}, x_i \right) = 0,\,\,  a = 1, 2\,\, ,
\end{equation}

\noindent
with unknown values of energy density $\epsilon_{1i}$, pressure $p_{1i}$,
velocity $v_{1i}$, and at the coordinate point $x_i$. Now let us consider the
evolution of the discontinuity between fluid and the gas of free particles.
First, let us count the unknowns. They are
$T_1$, $v_\sigma^{ref}$, $v_{1}$, $v_2$
(instead of $v_{rel}$) and $x^{shock}_i$.
The last one we can change to the velocity of the shock in the
laboratory system:
$v^{Lab}_\sigma$. Thus, we have 5 unknowns. On the other hand we have
precisely 5 equations. They are: two hydrodynamical equations for the fluid,
two conservation laws on the discontinuity between fluid and gas and
the relation between the velocities in different frames:

\begin{equation}
v_{\sigma}^{Lab}= \frac{v_{\sigma}^{ref} + v_{2}}{1 + v_{\sigma}^{ref} v_{2}}
\,\, .
\end{equation}

\noindent
Thus, solving this system of equations one finds the trajectory of the
freeze-out hypersurface, temperature of the fluid and velocities of fluid
and gas.

\vspace{0.2cm}

\noindent
{\bf B.} Now let us consider the case of baryonic matter. Then one has
more unknowns and more equations. Let us count them.

The unknown quantities are: three thermodynamical functions of the fluid -
$\epsilon_1$, $p_1$, $n_1$,
two thermodynamical functions for the gas - $T_2$, $\mu_2$,
the  derivative to the freeze-out hypersurface in
the reference frame of the gas - $v_\sigma^{ref}$,
the velocities of the fluid and gas in the laboratory system $v_1$, $v_2$,
and the derivative to the freeze-out hypersurface in the laboratory system
$v_{\sigma}^{Lab}$. Thus, one has 9 unknowns. At the same time there are
also 9 equations:
3 hydrodynamical equations for the fluid (two for energy and momentum, one
for the baryonic charge), 3 equations on the discontinuity between fluid and
gas (two for energy and momentum, one for the baryonic charge),
the equation of state for the fluid that relates its thermodynamical functions
$\epsilon_1$, $p_1$, $n_1$, the relation between the
velocity of the shock in the reference frame
of the gas and in the laboratory frame,
and the  equation of freeze-out for the gas of free
particles $n_2(x,t) = n^*$. Now one has to solve the above system of equations
and find all unknowns.

\newpage
\noindent
\begin{picture}(0,0)%
\includegraphics{dec1.pstex}%
\end{picture}%
\setlength{\unitlength}{0.002500in}%
\begingroup\makeatletter\ifx\SetFigFont\undefined
\def\x#1#2#3#4#5#6#7\relax{\def\x{#1#2#3#4#5#6}}%
\expandafter\x\fmtname xxxxxx\relax \def\y{splain}%
\ifx\x\y   
\gdef\SetFigFont#1#2#3{%
  \ifnum #1<17\tiny\else \ifnum #1<20\small\else
  \ifnum #1<24\normalsize\else \ifnum #1<29\large\else
  \ifnum #1<34\Large\else \ifnum #1<41\LARGE\else
     \huge\fi\fi\fi\fi\fi\fi
  \csname #3\endcsname}%
\else
\gdef\SetFigFont#1#2#3{\begingroup
  \count@#1\relax \ifnum 25<\count@\count@25\fi
  \def\x{\endgroup\@setsize\SetFigFont{#2pt}}%
  \expandafter\x
    \csname \romannumeral\the\count@ pt\expandafter\endcsname
    \csname @\romannumeral\the\count@ pt\endcsname
  \csname #3\endcsname}%
\fi
\fi\endgroup
\begin{picture}(490,495)(110,290)
\end{picture}

\newpage
\noindent
\begin{picture}(0,0)%
\includegraphics{dec2.pstex}%
\end{picture}%
\setlength{\unitlength}{0.002500in}%
\begingroup\makeatletter\ifx\SetFigFont\undefined
\def\x#1#2#3#4#5#6#7\relax{\def\x{#1#2#3#4#5#6}}%
\expandafter\x\fmtname xxxxxx\relax \def\y{splain}%
\ifx\x\y   
\gdef\SetFigFont#1#2#3{%
  \ifnum #1<17\tiny\else \ifnum #1<20\small\else
  \ifnum #1<24\normalsize\else \ifnum #1<29\large\else
  \ifnum #1<34\Large\else \ifnum #1<41\LARGE\else
     \huge\fi\fi\fi\fi\fi\fi
  \csname #3\endcsname}%
\else
\gdef\SetFigFont#1#2#3{\begingroup
  \count@#1\relax \ifnum 25<\count@\count@25\fi
  \def\x{\endgroup\@setsize\SetFigFont{#2pt}}%
  \expandafter\x
    \csname \romannumeral\the\count@ pt\expandafter\endcsname
    \csname @\romannumeral\the\count@ pt\endcsname
  \csname #3\endcsname}%
\fi
\fi\endgroup
\begin{picture}(490,495)(110,290)
\end{picture}

\end{document}